# In-plane plasmonic modes in a quasicrystalline array of metal nanoparticles

Zi-Lan Deng[*], Jian-Wen Dong[*], and He-Zhou Wang[**]

State Key Laboratory of Optoelectronic Materials and Technologies, Sun Yat-Sen (Zhongshan) University, Guangzhou 510275, China

## Abstract

We investigated the plasmonic modes in a two-dimensional quasicrystalline array of metal nanoparticles. The polarization of the modes is in the array plane. A simplified eigen-decomposition method is presented with the help of rotational symmetry. Two kinds of anti-phase ring modes with radial and tangential polarizations are of highest spatial localizations among all of plasmonic modes. For the leaky characteristic of the anti-phase ring modes, the highest fidelity mode is found to be tangential polarized mode in the quasicrystal array, whereas normal-to-plane polarized mode in the solo ring. The leaky characteristics and spatial localizations of other plasmonic modes are also studied, for example, collective vortex mode that may be potentially useful to form negative responses in plasmonic device, and collective radial mode that may be used to generate light sources with radial polarizations.

**Keywords:** surface plasmon polariton, metal nanoparticle, quasicrystal, localization

* Corresponding author: dongjwen@mail.sysu.edu.cn, 8620-8411-1469

** Corresponding author: stswhz@mail.sysu.edu.cn

# Introduction

Plasmonic materials have been paid tremendous attentions due to the significant improvement in fabrication techniques [1,2] and their practical applications. The subwavelength phenomena and strong near-field enhancement near optical frequencies [3,4], enable us to use plasmonic materials as building blocks for a variety of nanoscale optical applications [5], such as biosensors [6,7], subwavelength waveguides [8,9], optical left-handed metamaterials [10], plasmonic and hybrid waveguides[11, 12], invisibility[13], beam splitter[14] and even plasmonic sources [15].

Besides applications, plasmonic systems are interested because the underlying physics is somewhere in between electronic and photonic system, but has unique academic interest [16-21]. The electronic system has true eigenmodes derived from bound states. The localization is entirely concerned with the spatial localization of the electronic wavefunctions [22-24]. For two-dimensional dielectric photonic systems, it also has true eigenstates derived from scattering states. The localization is still an essential study of spatial properties in real space eigenmodes [23,24]. Although plasmonic systems are intrinsic long-range in coupling similar to photonic crystal, the resonances bear resemblance to the electronic bound states, more than to the photonic scattering states. For two-dimensional plasmonic system, it has radiation loss and thus we are dealing with quasi-eigenmodes. An extra dimension should be added in order to describe the fidelity or the radiation loss of the localized modes. Thus, we are dealing with the localization properties of "leaky" states with long range couplings in an open system.

Up to now, Anderson localization of plasmonic eigenmodes in random arrays [19-21], as well as leaky wave of quasicrystalline structures [18,25], has been studied. The physics is anticipated to be richer and more complicated than those in electronic and photonic systems. In this paper, the characteristics of plasmonic modes with the polarization in the array plane, i.e. in-plane (IP) modes, are investigated with the help of rotational symmetry analysis and eigen-decomposition method. The resonant frequency spectra of radiating and weakly-radiating in-plane modes are found to mix together, which is more complicated than the normal-to-plane (NP) modes (polarization is perpendicular to array plane) studied previously [25]. The plasmonic mode with highest spatial localization and high fidelity is found to be a type of in-plane ring mode with tangential polarization, which is different from the solo ring case [26]. In addition, different localized modes have very different radial decay behaviors and there is no special relationship between the fidelity of the modes and their spatial localization.

## Eigen-decomposition method in quasicrystalline system

The plasmonic system under study is a metal nanoparticles (MNP) array arranged as two-dimensional (2D) quasicrystal (QC) with 12-fold symmetry generated by the generalized dual method [27]. We assume each particle has a dynamic dipole polarizability, $\alpha = i(3/2k_0^3)a_1$, where $k_0 = \omega/c$ with $c$ as the speed of light, and $a_1$ is the electric term of the Mie's coefficients [28]. We use Drude-type permittivity $\varepsilon(\omega) = 1 - \omega_p^2/(\omega^2 + i\gamma\omega)$ with $\omega$ as the angular frequency, $\omega_p$ as the plasma frequency, and $\gamma$ as the electron scattering rate. In order to meet the criteria of the dipole approximation, we restrict ourselves to the situations when the particles are not too close together with a condition of $a_{min} \geq 3r_0$, where $a_{min}$ is the distance between two nearest particles and $r_0$ is the radius of the particle. Here we set $\omega_p = 6.18 eV$, $a_{min} = 75 nm$, $r_0 = 25 nm$, $\gamma = 0$.

We assume that an external time-harmonic driving electric field $\mathbf{E}_m^{ext} e^{-i\omega t}$ acts on the $m$-th particle located at $\mathbf{R}_m$, and the coupled dipole equations can be written as the form,

$$\alpha^{-1}\mathbf{p}_m - \sum_{m' \neq m} \overset{\leftrightarrow}{\mathbf{W}}(\mathbf{R}_m - \mathbf{R}_{m'})\mathbf{p}_{m'} = \mathbf{E}_m^{ext}, \qquad (1)$$

where $\mathbf{p}_m \equiv (\mathbf{p}_{m,//}, p_{m,z})$ is the $m$-th dipole moment, and $\overset{\leftrightarrow}{\mathbf{W}}(\cdot)$ is the dynamic Green's function. Eq. (1) can be written in a matrix form as $\mathbf{Mp} = \mathbf{E}$, where $\mathbf{M} = \alpha^{-1}\mathbf{I} - \sum \overset{\leftrightarrow}{\mathbf{W}}$. For a 2D system, we can divide the equation as $\mathbf{M}_{//}\mathbf{p}_{//} = \mathbf{E}_{//}$ for IP modes, and $\mathbf{M}_{zz}\mathbf{p}_z = \mathbf{E}_z$ for NP modes. We apply the eigen-decomposition (ED) method [16,17] to analyze the electromagnetic resonances of such a plasmonic system. This method is more efficient than the previous ones that need either numerical complex root searching [29] or root approximation [30] in complex plane. In ED method, the eigenpolarizability with the form of $\alpha_{eig} = 1/\lambda$ can be well defined, where $\lambda$ is the complex eigenvalues of the matrix $\mathbf{M}$. This quantity enables us to describe the collective resonance strength of the whole system for an external electric field pattern that is proportional to the corresponding eigenmode. Moreover, the participation ratio (PR) with the form of $P_n = \left(\sum_{m=1}^{N}|\hat{\mathbf{p}}_m^{(n)}|^2\right)^2 \bigg/ \left(\sum_{m=1}^{N}|\hat{\mathbf{p}}_m^{(n)}|^4\right)$ is a good indicator to measure the order of the spatial localization of the modes, where $\hat{\mathbf{p}}_m^{(n)}$ is the n-th complex eigenvectors of the matrix $\mathbf{M}$. The smaller value of $P_n$, the higher spatial localization [25].

Now we classify the eigenmodes by using rotational symmetry of the $F$-fold plasmonic system. To simplify the eigenvalue problem, we divide the lattice points into $M+1$ sets, where $M \equiv (N-1)/F$ and $N$ is the total number of particles. So each set (except the one which contains only the central point located at the origin) contains $F$ lattice points that equidistantly lie on a single ring. Then we transform Cartesian coordinates to polar coordinates, i.e. $\mathbf{p'} = \mathbf{\Omega p}$ where $\mathbf{\Omega}$ is the rotational matrix. After the transform, the eigenvalue problem becomes $\mathbf{M'p'} = \lambda \mathbf{p'}$, where $\mathbf{M'} = \mathbf{\Omega M \Omega}^{-1}$. Due to the rotational symmetry, $\mathbf{M'}$ remains invariant under cyclic index transform, i.e., $\mathbf{TM'T^{-1}} = \mathbf{M'}$, where

$$\mathbf{T} = \begin{pmatrix} \tilde{\mathbf{I}} & \mathbf{O} & \mathbf{O} & \mathbf{O} & \mathbf{O} \\ \mathbf{O} & \tilde{\mathbf{T}} & \mathbf{O} & \mathbf{O} & \mathbf{O} \\ \mathbf{O} & \mathbf{O} & \tilde{\mathbf{T}} & \mathbf{O} & \mathbf{O} \\ \mathbf{O} & \mathbf{O} & \mathbf{O} & \ddots & \vdots \\ \mathbf{O} & \mathbf{O} & \mathbf{O} & \cdots & \tilde{\mathbf{T}} \end{pmatrix}, \text{ and } \tilde{\mathbf{T}} = \begin{pmatrix} \mathbf{O} & \tilde{\mathbf{I}} & \mathbf{O} & \mathbf{O} & \mathbf{O} \\ \mathbf{O} & \mathbf{O} & \tilde{\mathbf{I}} & \ddots & \vdots \\ \mathbf{O} & \mathbf{O} & \mathbf{O} & \ddots & \mathbf{O} \\ \mathbf{O} & \mathbf{O} & \mathbf{O} & \ddots & \tilde{\mathbf{I}} \\ \tilde{\mathbf{I}} & \mathbf{O} & \mathbf{O} & \cdots & \mathbf{O} \end{pmatrix}, \quad (2)$$

where $\tilde{\mathbf{I}}$ is a 3×3 identity matrix. The eigenvector of cyclic matrix $\tilde{\mathbf{T}}$ can be written as

$$\tilde{R}_{ju}^{(J)} = c_u^{(J)} e^{i2\pi Jj/F}, \quad (3)$$

where $c_u^{(J)}$ is an arbitrary constant, $J, j = 1, 2, ..., F$, and $u = 1, 2, 3$. Since $\mathbf{T}$ is formed by combining $\tilde{\mathbf{I}}$ and $\tilde{\mathbf{T}}$ as the block-diagonal elements, the eigenvectors of $\mathbf{T}$ have the form of,

$$\begin{cases} \left(R_{0u}^{(J)}, R_{1ju}^{(J)}, R_{2ju}^{(J)}, ..., R_{Mju}^{(J)}\right) = \left(q_{0u}^{(J)}, q_{1u}^{(J)} e^{i2\pi Jj/F}, q_{2u}^{(J)} e^{i2\pi Jj/F}, ..., q_{Mu}^{(J)} e^{i2\pi Jj/F}\right), & J = F \\ \left(R_{1ju}^{(J)}, R_{2ju}^{(J)}, ..., R_{Mju}^{(J)}\right) = \left(q_{1u}^{(J)} e^{i2\pi Jj/F}, q_{2u}^{(J)} e^{i2\pi Jj/F}, ..., q_{Mu}^{(J)} e^{i2\pi Jj/F}\right), & J \neq F \end{cases}, \quad (4)$$

where $q_{ku}^{(J)}$ are arbitrary constants. Since $\mathbf{M}'$ and $\mathbf{T}$ commute, they possess the same eigenvectors, and thus, we can write the eigenvectors of $\mathbf{M}'$ in the form of,

$$\begin{cases} \left(p'^{(J,K,U)}_{0u}, p'^{(J,K,U)}_{1ju}, ..., p'^{(J,K,U)}_{Mju}\right) = \left(q_{0u}^{(J,K,U)}, q_{1u}^{(J,K,U)} e^{i2\pi Jj/F}, ..., q_{Mu}^{(J,K,U)} e^{i2\pi Jj/F}\right), & J = F \\ \left(p'^{(J,K,U)}_{1ju}, ..., p'^{(J,K,U)}_{Mju}\right) = \left(q_{1u}^{(J,K,U)} e^{i2\pi Jj/F}, ..., q_{Mu}^{(J,K,U)} e^{i2\pi Jj/F}\right), & J \neq F \end{cases}, \quad (5)$$

where $q_{ku}^{(J,K,U)}$ are another set of arbitrary constants that is different from those in Eq. (4). Here $U = 1, 2, 3$; $K, k = 0, 1, 2, \cdots, M$ for $J = F$ and $K, k = 1, 2, \cdots, M$ for $J \neq F$. In Eqs. (3)-(5), the superscripts $(K, J, U)$ are the mode indices, and the subscripts $(k, j, u)$ are the vector element indices. Taking (5) into $\mathbf{M}'\mathbf{p}' = \lambda \mathbf{p}'$, we obtain

$$\begin{cases} \sum_{k'=0}^{M} \sum_{j'=1}^{F} \sum_{u'=1}^{3} \mathbf{M}'_{kk'jj'uu'} q_{k'u'}^{(J,K,U)} = \lambda^{(J,K,U)} q_{ku}^{(J,K,U)}, & J = F \\ \sum_{k'=1}^{M} \sum_{j'=1}^{F} \sum_{u'=1}^{3} \mathbf{M}'_{kk'jj'uu'} q_{k'u'}^{(J,K,U)} e^{i2\pi Jj'/F} = \lambda^{(J,K,U)} q_{ku}^{(J,K,U)} e^{i2\pi Jj/F}, & J \neq F \end{cases} \quad (6)$$

In order to eliminate $e^{i2\pi Jj/F}$ in Eq. (6), we set $j=F$, and exchange the summation order of $j'$ and $u'$, yielding,

$$\begin{cases} \sum_{k'=0}^{M} \sum_{u'=1}^{3} \left( \sum_{j'=1}^{F} \mathbf{M}'_{kk'Fj'uu'} \right) q_{k'u'}^{(J,K,U)} = \lambda^{(J,K,U)} q_{ku}^{(J,K,U)}, & J = F \\ \sum_{k'=1}^{M} \sum_{u'=1}^{3} \left( \sum_{j'=1}^{F} \mathbf{M}'_{kk'Fj'uu'} e^{i2\pi Jj'/F} \right) q_{k'u'}^{(J,K,U)} = \lambda^{(J,K,U)} q_{ku}^{(J,K,U)}, & J \neq F \end{cases} \quad (7)$$

Introducing a new matrix

$$\tilde{\mathbf{M}}'^{(J)}_{kk'uu'} = \sum_{j'=1}^{F} \mathbf{M}'_{kk'Fj'uu'} e^{i2\pi Jj'/F} \quad \left(u,u'=1,2,3;\ k,k' = \begin{cases} 0,1,...,M \text{ for } J=F \\ 1,2,...,M \text{ for } J \neq F \end{cases}\right). \tag{8}$$

The eigenvalue problem becomes $\tilde{\mathbf{M}}'^{(J)}\mathbf{q}^{(J)} = \lambda^{(J)}\mathbf{q}^{(J)}$. We note that the dimension of $\tilde{\mathbf{M}}'^{(J)}$ is $3(M+1) \times 3(M+1)$ for $J=F$, and $3M \times 3M$ for $J \neq F$. By exchanging index $J$ to $F-J$, and $j'$ to $F-j'$ in Eq. (8), we have $\tilde{\mathbf{M}}'^{(F-J)} = \tilde{\mathbf{M}}'^{(J)}$. It indicates that the eigenmodes indexed by $J$ and $F-J$ are degenerate. For 12-fold QC system ($F=12$), the $J=1,2,3,4,5$ eigenmodes are degenerate with the $J=11,10,9,8,7$ eigenmodes, respectively; while the eigenmodes with $J=12$ (in-phase) and $J=6$ (anti-phase) are in general non-degenerate. So there are seven kinds of modes with different rotational symmetry. In the calculations, we divide $\tilde{\mathbf{M}}'^{(J)}$ into $\tilde{\mathbf{M}}'^{(J)}_{//}$ and $\tilde{\mathbf{M}}'^{(J)}_{zz}$ in order to obtain IP and NP plasmonic modes, respectively. We use standard numerical routines in LAPACK to diagonalize $\tilde{\mathbf{M}}'^{(J)}_{//}$ and $\tilde{\mathbf{M}}'^{(J)}_{zz}$ for each $J$. With the help of the above method, the dimension of the matrix is significantly reduced so that the plasmonic modes of large sample (e.g. N ~ 10000) can be solved out. This is impossible if we diagonalize the matrix $\mathbf{M}$ directly with limited computation resources and acceptable compuation time. Unfortunely, the sequence of the eigenmodes is random, and it seems hard to connect the eigenvalues from one frequency to another. Thanks to the symmetry of the QC, we use the similarity of the coefficients $q_{ku}^{(J,K,U)}$ between neighboring frequencies to connect eigenvalues readily. Note that this connection cannot be done if we diagonalize the matrix $\mathbf{M}$ directly.

## Results and Discussions

Figure 1 shows an intensity plot of $\text{Im}(\alpha_{eig})$ on the plane of frequency and mode index in the 12-fold QC with $N=1513$. Here we use the symbol $m$ instead of $(K,U)$ to represent the mode index. Figs. 1(a) and 1(b) are the cases of the anti-phase($J=6$) NP and in-phase($J=12$) NP modes; while Figs. 1(c) and 1(d) are the cases of the anti-phase IP and in-phase IP modes, respectively. Here we only show the results of $J=6$ and $J=12$ since the modes with other $J$ indices are quite similar. As a spectral function, the peaks of $\text{Im}(\alpha_{eig})$ represent resonances. The mode index, m, is used to identify the mode whose resonant frequency is sorted in an increasing order. In fact, $\text{Im}(\alpha_{eig})$ is proportional to the extinction of the driving field, and the width of the peak in the spectra shows automatically the mode quality (fidelity). The larger the peak width is, the less dramatic color changes, and the stronger radiating the mode is. It can be seen, for the NP modes, that the spectral function has narrow and conspicuous peaks near 3.25 eV, while large and smooth peaks located in lower and higher frequency ranges. It indicates that the modes with weakly out-of-plane radiating loss are separated from the radiating modes at both sides. This is consistent with our previous findings [25]. But this explicit characteristic is not found in the spectra of IP modes. It can be seen, from Figs. 1(c) and 1(d), that the radiating and weakly-radiating modes are mixed together in the frequency range.

In Ref. [25], we introduced a trajectory map in a three-dimensional parameter space $(\omega,|\lambda|,P_n)$, which can give information about the temporal properties and the spatial properties of the modes. In this paper, we follow the idea but use a more simple representation. We pick out the resonant frequency for each mode, and dot it in a two-dimensional parameter space $\left[ \text{Im}(\alpha_{eig}), P_n \right]$. Figure 2 shows the results for different kinds of anti-phase IP plasmonic modes in different sizes of the MNP clusters. On one hand, there are localized modes, whose PR values are much smaller than the cluster number, and they almost remain the same when the size of the cluster increases. On the other hand, there are also extended modes whose PR values are expanded to approximately twice when the particles of the cluster increase from 1513 to 3529. Figure 2 also shows that there is no special relationship between the radiation (fidelity) of the modes and their spatial localization. What's more, it is interesting to find that there are two anti-phase IP modes with the largest $\text{Im}(\alpha_{eig})$ and the smallest PR in the bottom-right of Figs. 2(a) and 2(b). It indicates that they are highest localized and have highest fidelity. As discussed below, these two IP modes are the radial anti-phase ring (RAPR) mode, and the tangential anti-phase ring (TAPR) mode.

Next, we will turn to focus on two anti-phase ring modes. Figs. 3(a) and 3(b) are the mode patterns of RAPR, and TAPR modes, respectively. In order to compare with the ring mode whose dipole moment directions point to normal-to-plane, we also plot the mode pattern of normal anti-phase ring (NAPR) mode in Fig. 3(c). Because the excited dipole moments of these three ring modes are strongly confined on the central ring of the QC, we just show the dipole moments near the center. The magnitude of the dipole is exactly the same in every particle within the central ring. But the direction of the dipole is different from each mode. It is along radial direction of array plane for the RAPR mode, along tangential direction of array plane for the TAPR mode, and perpendicular to the array plane for the NAPR mode. In addition, no matter which anti-phase ring mode in Figs. 3(a)-3(c), the direction of the neighboring dipoles are opposite in sign. As shown in Fig. 2, we find that the anti-phase ring modes are highest spatially localized and have highest fidelity. In order to give more evidence, we have calculated for larger size of clusters, and the results are shown in Figs. 3(d)-3(f). It is found that the resonant frequency and the PR value are almost unchanged even when the number of particles reaches over 10,000. It indicates that the anti-phase ring modes are indeed well-define localized modes. Moreover, the resonant frequency of both NAPR and RAPR are near the resonant frequency of a single particle (3.25eV), while the resonant frequency of TAPR (3.63eV) is largely above that of a single particle [see Fig. 3(d)]. This is analogous to the ring modes in a circular array [26]. Fig. 3(f) shows that the NAPR mode has the highest fidelity when the cluster has only a ring pattern (N=13), which is in accordance with Ref. [26]. However, when the cluster arranges into the QC shape, simulation results show that the mode with highest fidelity is the TAPR mode instead of the NAPR mode.

Apart from the anti-phase ring modes, there are many other different localized modes. For example, there is a kind of in-phase ring IP mode, of which the energy is concentrated on the inner two rings near the center point. All dipoles are radially polarized and in-phase [see Fig. (4a)]. From its Fourier spectrum, it is found that the in-phase IP mode is strong radiating and has low mode quality due to the boundary match to freespace. There is another typical localized mode with three-fold rotational symmetry, whose mode pattern is shown in Fig. 4(b). The dipole moments concentrate on a large ring of the QC. The PR value of such large ring mode is about 42, much small than the particle number 1513. In addition, we find some plasmonic modes with collective vortex patterns and collective radial patterns. Two representative examples are shown in Figs. 4(c) and 4(d). The dipole moments are localized in the central area of QC array, and point to azimuthal direction and radial direction in phase. The vortex mode is promising to induce a large magnetic field so that it may be used to form negative response and plasmonic metamaterials. The radial mode may be potentially used to produce light sources with radial polarizations [31].

## Summary

In summary, we have investigated the rotational symmetry, the radiating and the spatial localization properties of in-plane plasmonic eigenmodes in the QC MNP lattice. A simplified form of the eigen-problem has been derived, so that the overwhelming information in the large system can be reduced according to the symmetry of the QC. Two anti-phase ring modes with different polarizations are found to be of high fidelity and high spatial localization. In particular, the anti-phase ring mode with tangential polarization has the highest fidelity. Other localized mode and the collective vortex and collective radial mode are also studied.

## Acknowledgments

We would like to thank Prof. C. T. Chan for his valuable discussions. This work is supported by the National Natural Science Foundation of China (10804131, 11074311, 10874250), the Fundamental Research Funds for the Central Universities (2009300003161450), and the Guangdong Natural Science Foundation (10451027501005073).

**Figure captions**

FIG. 1. (Color online) Intensity plot of $\log_{10}\left[\text{Im}(\alpha_{eig})\right]$ for the resonant spectra of plasmonic eignmodes with a 12-fold quasicrystal. (a) and (b) for the anti-phase and in-phase normal-to-plane modes, while (c) and (d) for the anti-phase and in-phase in-plane modes, respectively.

FIG. 2. (Color online) Participation ratio v.s. $\text{Im}(\alpha_{eig})$ at resonant frequency for the anti-phase in-plane modes with different numbers of particles. The particle number of the cluster is 1513 and 3529, respectively.

FIG. 3. (Color online) Characteristics of three anti-phase ring modes with different number of particles. The dipole distributions for the (a) RAPR, (b) TAPR, and (d) NAPR mode. The directions and lengths of arrows represents polarization directions and magnitudes, and the red/blue color represents positive/negative maximum. (d) The resonant frequency, (e) the PR value, and (f) $\text{Im}(\alpha_{eig})$ as the function of the number of particles.

FIG. 4. The dipole distributions of other different type of localized modes. (a) in-phase ring in-plane mode, (b) in-plane large ring mode, (c) collective vortex mode, and (d) collective radial mode. number of particles of the cluster is N=1513.

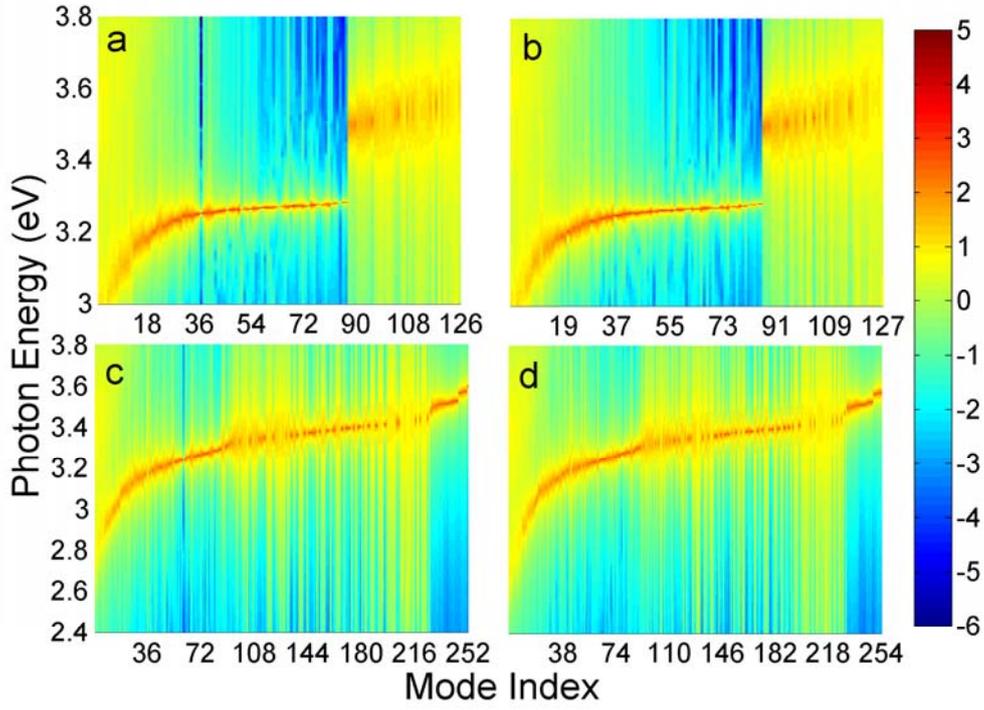

**FIG. 1. (Color online) Intensity plot of** $\log_{10}\left[\text{Im}(\alpha_{eig})\right]$ **for the resonant spectra of plasmonic eignmodes with a 12-fold quasicrystal. (a) and (b) for the anti-phase and in-phase normal-to-plane modes, while (c) and (d) for the anti-phase and in-phase in-plane modes, respectively.**

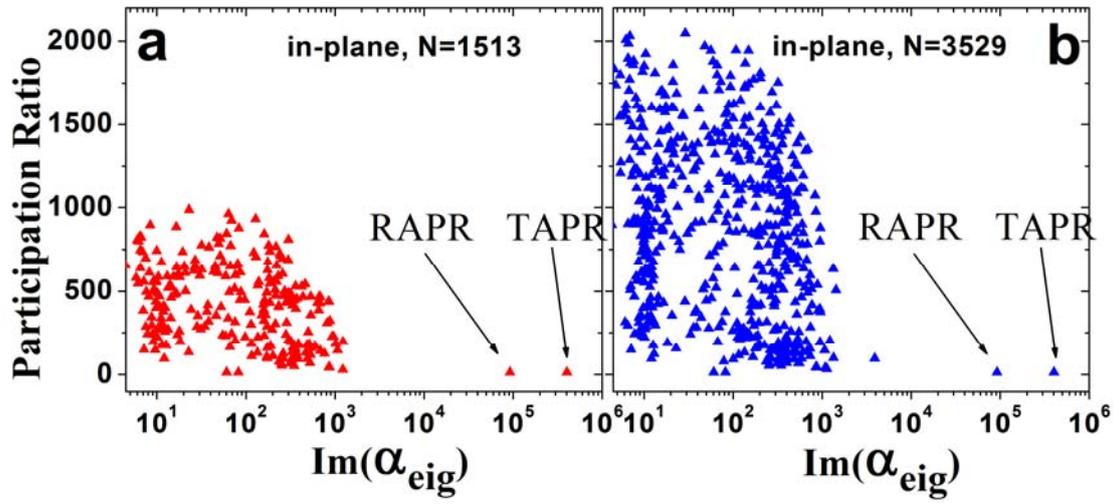

**FIG. 2. (Color online)** Participation ratio v.s. $\text{Im}(\alpha_{eig})$ at resonant frequency for the anti-phase in-plane modes with different numbers of particles. The particle number of the cluster is 1513 and 3529, respectively.

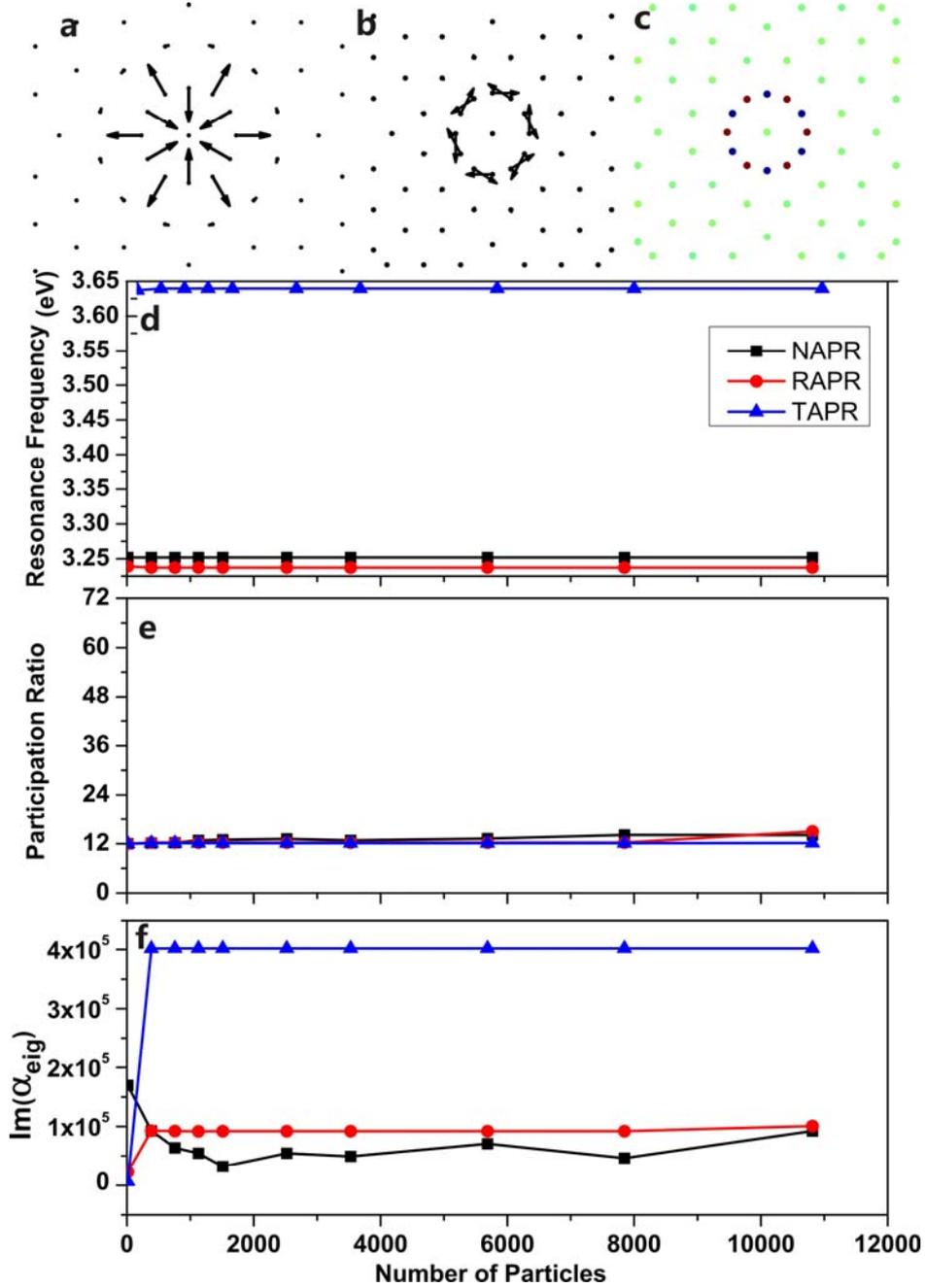

FIG. 3. (Color online) Characteristics of three anti-phase ring modes with different number of particles. The dipole distributions for the (a) RAPR, (b) TAPR, and (d) NAPR mode. The directions and lengths of arrows represents polarization directions and magnitudes, and the red/blue color represents positive/negative maximum. (d) The resonant frequency, (e) the PR value, and (f) $\mathrm{Im}(\alpha_{eig})$ as the function of the number of particles.

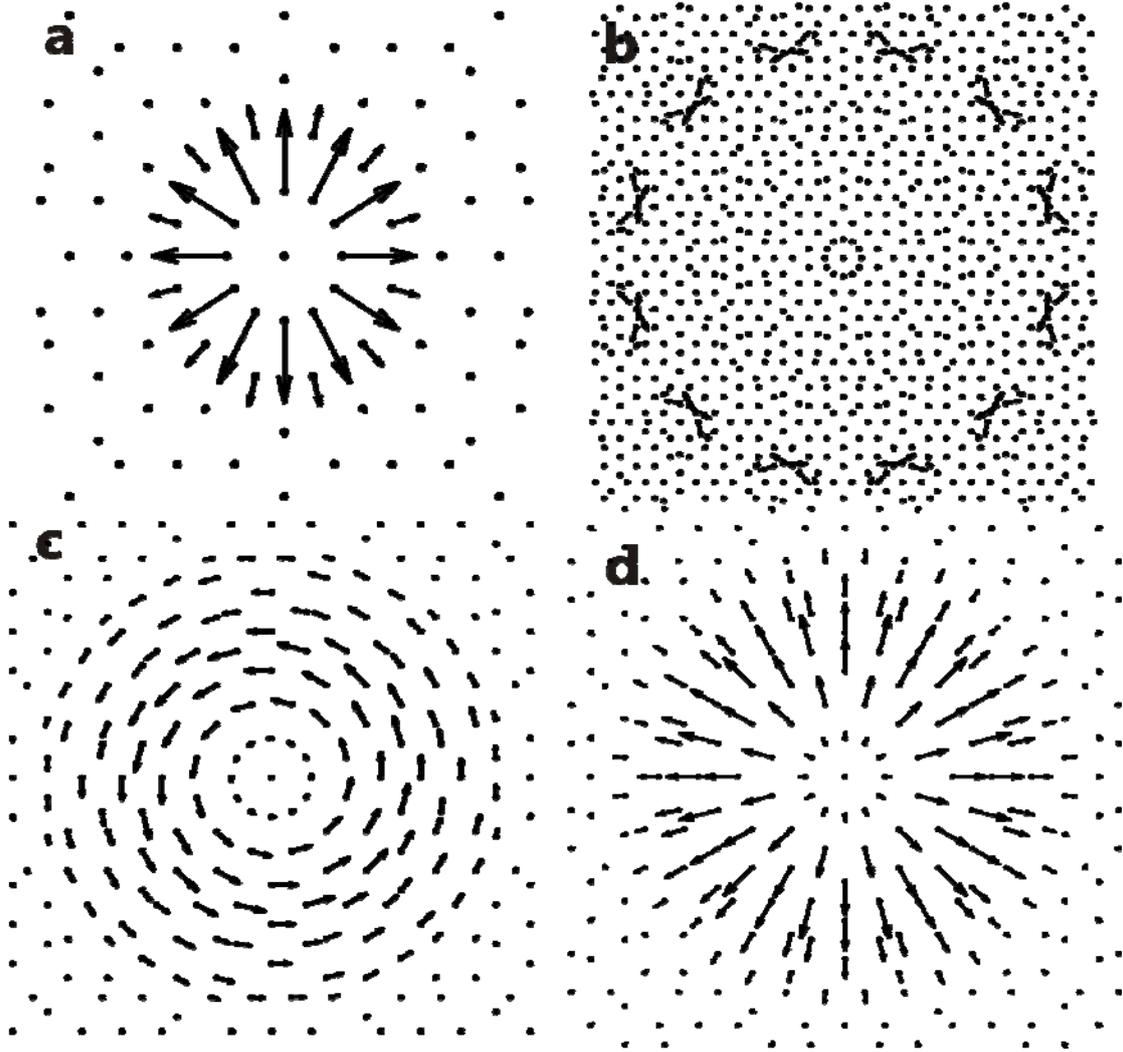

FIG. 4. The dipole distributions of other different type of localized modes. (a) in-phase ring in-plane mode, (b) in-plane large ring mode, (c) collective vortex mode, and (d) collective radial mode. number of particles of the cluster is N=1513.